\documentclass[aps,prb,twocolumn,groupedaddress,draft,showpacs,floatfix]{revtex4}
\usepackage{graphicx}

\begin{document}

\newcommand{\be}{\begin{equation}}
\newcommand{\ee}{\end{equation}}
\newcommand{\bea}{\begin{eqnarray}}
\newcommand{\eea}{\end{eqnarray}}
\newcommand{\bi}{\bibitem}
\newcommand{\da}{\downarrow}
\newcommand{\la}{\langle}
\newcommand{\ra}{\rangle}

\title{Impurity and boundary effects in one and two-dimensional 
inhomogeneous Heisenberg antiferromagnets}
\author{P. E. G. Assis, Valter L. L\'{\i}bero and K. Capelle}
\affiliation{Departamento de F\'{\i}sica e Inform\'atica,
Instituto de F\'{\i}sica de S\~ao Carlos,
Universidade de S\~ao Paulo,\\
Caixa Postal 369, 13560-970 S\~ao Carlos, SP, Brazil}
\date{\today}

\begin{abstract}
We calculate the ground-state energy of one and two-dimensional spatially
inhomogeneous antiferromagnetic Heisenberg models for spins $1/2$, $1$, $3/2$ 
and $2$. Our calculations become possible as a consequence of the recent 
formulation of density-functional theory for Heisenberg models. The method 
is similar to spin-density-functional theory, but employs a local-density-type 
approximation designed specifically for the Heisenberg model, allowing us to 
explore parameter regimes that are hard to access by traditional methods, and 
to consider complications that are important specifically for nanomagnetic 
devices, such as the effects of impurities, finite-size, and boundary 
geometry, in chains, ladders, and higher-dimensional systems. 
\end{abstract}



\maketitle

The study of low-dimensional spin systems is one of the central issues in
the physics of correlated electrons. Model Hamiltonians of the Heisenberg
type are widely used to study, e.g., antiferromagnetically coupled spin chains,
ladders, and layers,\cite{dagotto,fulde} and constitute most useful models
for strong correlations in undoped cuprates and manganites. Finite-size 
Heisenberg models, moreover, serve as paradigmatic models for the emerging
field of nanomagnetism and spintronics. However, progress in the analysis
of nanomagnetic devices requires capability to deal with `real-life' 
complications, such as impurities of arbitrary size and location, boundaries 
of various shapes, and crossovers between finite and extended systems, or 
between one and higher dimensions. In the present paper we present a convenient
and efficient numerical approach for calculating ground-state energies of
antiferromagnetic Heisenberg models subjected to such complications.

Traditional numerical methods are well suited for studying generic properties
of homogeneous Heisenberg models, and have provided many important insights 
into the physics of magnetic systems. However, the very significant expenditure
of computational effort required by methods such as group-theory-aided exact 
diagonalization, Quantum Monte Carlo (QMC),\cite{qmc} or the density-matrix 
renormalization group (DMRG),\cite{lou} imposes rather strict limits 
on the size and complexity of systems that can be studied. 
The mean-field approximation can, in principle, be 
applied to systems of almost arbitrary size and complexity, but its neglect 
of correlation makes it unreliable as a tool for studying many issues of 
current physical interest in strongly correlated systems.
In {\it ab initio} calculations, density-functional theory (DFT) provides a
convenient and reliable way to go beyond the mean-field (Hartree)
approximation, allowing to study real systems of considerable size and
complexity.\cite{kohnrmp,science,kubler} While DFT is in principle an exact 
reformulation of the many-body problem,\cite{dftbook} its practical 
implementation requires the use of approximations for the exchange-correlation 
energy. Among the most important such approximations is the local-density 
approximation (LDA), the essence of which is to use the exchange-correlation 
energy of the uniform electron liquid {\it locally} as an approximation to 
the one of the crystal or molecule of interest.\cite{kohnrmp,dftbook} The 
formal framework of DFT and the LDA is not limited to the {\it ab initio} 
case, but can be applied to a large class of model Hamiltonians, 
too.\cite{balda,mott,oxford,hemo,magyar}

In the present paper we study the ground-state energy of one and
two-dimensional antiferromagnetic systems with broken translational
symmetry, as a function of spin $S$ and system size $N$. Our choice of
symmetry-breaking terms is dictated by complications expected to arise in
real nanomagnetic devices, and includes boundaries of diverse geometries, 
and different types of impurities.
Our calculations are performed within DFT for the Heisenberg model,
\be
\hat{H} = J \sum_{ij} \hat{\bf S}_i \cdot \hat{\bf S}_j
 + \sum_i {\bf B}_i \cdot \hat{\bf S}_i,
\label{inhomheis}
\ee
where ${\bf B}_i$ is a symmetry-breaking magnetic field, 
coupled to the spins. The simplest local-spin approximation (LSA) for the
correlation energy  of Hamiltonian (\ref{inhomheis}) reads\cite{hemo}
\be
E_{c,d=1}^{LSA}[{\bf S}_i] = - 0.36338\, J \sum_i |{\bf S}_i|
\label{ldafunc1d}
\ee
in one dimension, and
\be
E_{c,d=2}^{LSA}[{\bf S}_i] = - 0.316\, J \sum_i |{\bf S}_i|
\label{ldafunc2d}
\ee
in two dimensions. More sophisticated approximations are discussed in
Ref.~\onlinecite{hemo}.
Here $|{\bf S}_i|$ is the modulus of the classical vector ${\bf S}_i$, the
ground-state expectation value of the local spin operator $\hat{\bf S}_i$.
The total-energy functional then takes the form
\be
E[{\bf S}_i] = J \sum_{i,j} {\bf S}_i \cdot {\bf S}_j + 
E_c^{LSA}[{\bf S}_i] + \sum_i {\bf B}_i \cdot {\bf S}_i,
\ee
where the first term on the right-hand side is the mean-field result,
the second the LSA correlation correction to it, and the last describes the
coupling to (possibly spatially varying) magnetic fields.
Below we set ${\bf B}_i=0$ because we will mostly be interested in
intrinsic inhomogeneities such as surfaces and impurities, but both
LSA functionals can be applied for nonzero ${\bf B}_i=0$ as well. 
Minimization of $E[{\bf S}_i]$ with respect to ${\bf S}_i$ yields the
ground-state energy $E_0$. Minimization of the first (mean-field) term on its 
own results in the familiar antiferromagnetic N\'eel state with alternating 
up and down spins. For open boundary conditions \cite{footnote1} 
in one dimension this state 
has energy $E_{0,d=1}^{MF}(S)= - J S^2 (N-1)$, while for a two dimensional
square lattice $E_{0,d=2}^{MF}(S)= - 2 J S^2 N (N-1)$, where $N$ is the number 
of sites along the chain and along one side of the square, respectively.
$E_c^{LSA}$ provides an additive correction to $E_0^{MF}$.

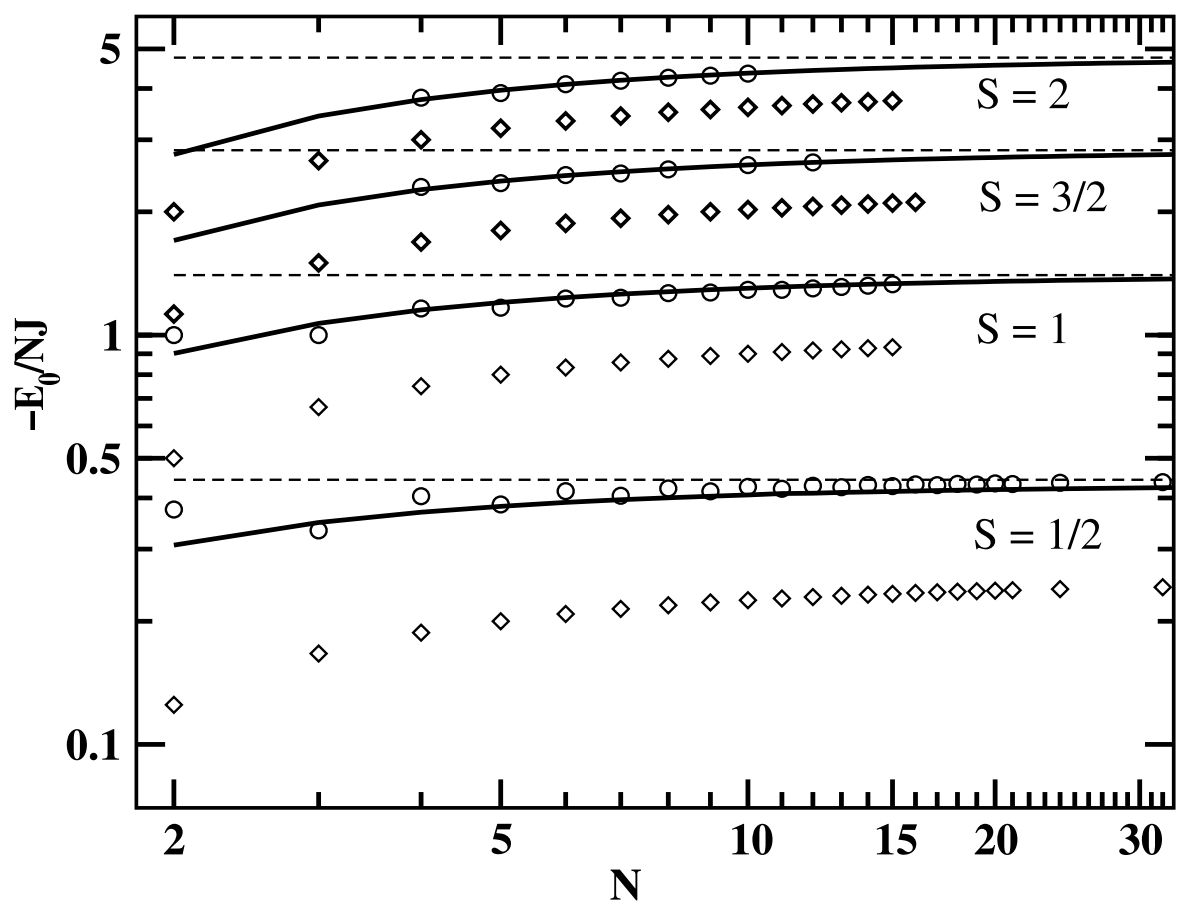
\begin{figure}
\caption{\label{fig1}Ground-state energy of the finite antiferromagnetic
Heisenberg chain for various values of the spin $S$.
For each spin, the diamonds represent the mean-field results, the circles
are numerically precise benchmark values obtained in independent many-body
calculations,\cite{footnote2} the full line is obtained with LSA, and the
dashed horizontal line represents the value obtained in the thermodynamic
limit, taken from Ref.~\onlinecite{lou}. The LSA is seem to provide
significant improvement on the mean-field values, for all $S$ and $N$.}
\end{figure}

First, we consider finite size Heisenberg chains. In Fig.~\ref{fig1} we 
compare, separately for each spin, the ground-state energies obtained in the 
mean-field approximation, in LSA, and by exact diagonalization. To be able 
to compare with exact results we have limited us to fairly small systems, 
but clearly mean-field and LSA calculations can be performed easily for much 
larger systems. The LSA is seen to provide substantial 
improvement on the mean-field results, at negligible extra computational 
cost. However, we also observe that the LSA does not reproduce the oscillatory
structure in the exact results visible for small $N$, but instead smoothly 
interpolates through it. As expected on physical grounds,\cite{hemo} the LSA 
becomes better as the system size increases.

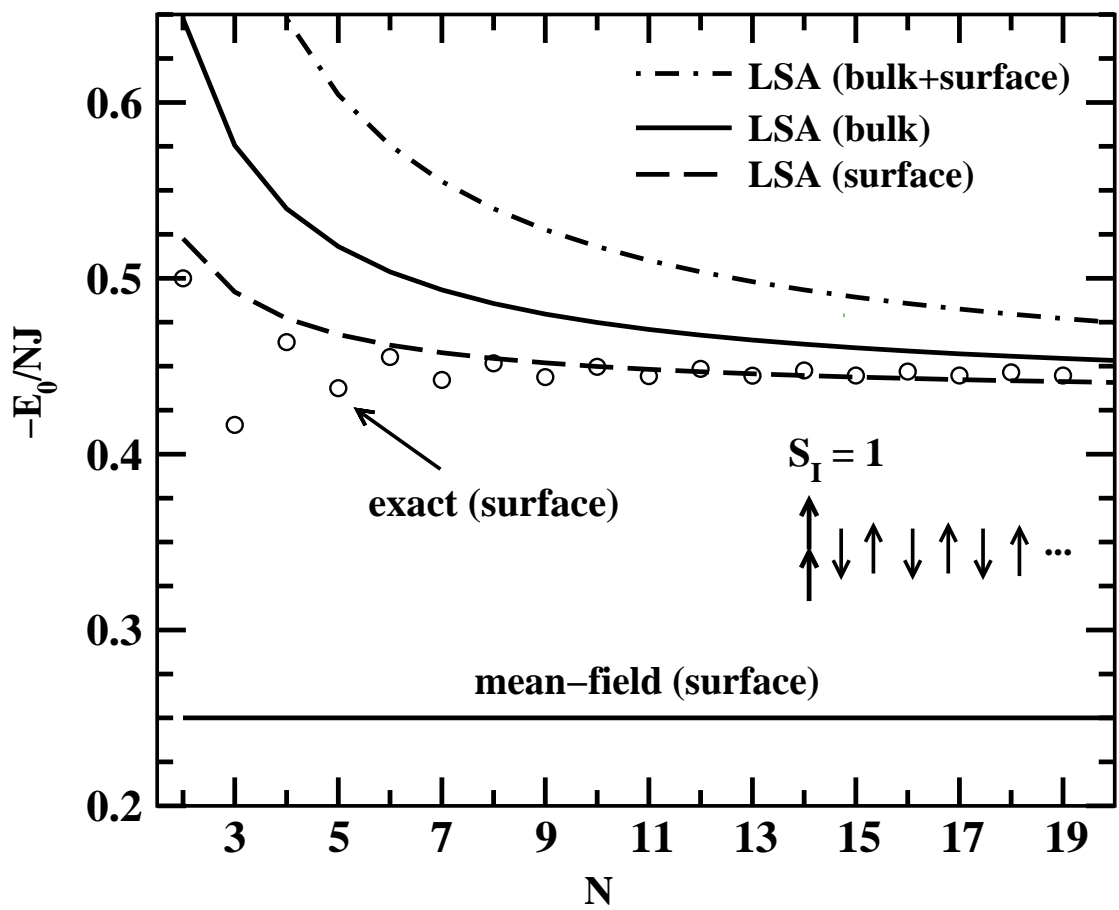
\begin{figure}
\caption{\label{fig2}Ground-state energy of an antiferromagnetic $S=1/2$
Heisenberg chain with an $S_I=1$ impurity at the surface. Horizontal line:
mean-field result. Dashed curve: LSA result. Circles: exact values, 
obtained by numerical diagonalization. Full curve: LSA result for 
same system, but with the impurity in the bulk. Dash-dotted curve: LSA
results for same system, but with two impurities.}
\end{figure}

In these calculations the system is inhomogeneous only due to its finite size. 
However, the LSA concept can also be applied to spin Hamiltonians in which 
translational symmetry is broken more radically. Specifically, we now consider 
an impurity model of much current physical interest: an impurity of spin 
$S_I$ in a magnetic host material of spin $S$. This type of model 
is relevant, e.g., for the analysis of impurities in the antiferromagnetic 
parent compounds of cuprate superconductors,\cite{dagotto} impurities in
correlated spin chains,\cite{kenzelmann} and recent proposals for quantum 
computing.\cite{castroneto,loss}
Fig.~\ref{fig2} shows the ground-state energy of a spin $1/2$ antiferromagnetic
Heisenberg chain, with open boundary conditions, in which one of the 
boundary spins has been replaced by a spin 1 ion, as schematically indicated
in the inset. Mean-field, LSA, and exact values are given. The exact data show
that for this type of impurity system the improvement on the mean-field
approximation provided by the LSA is of the same order as in the 
homogeneous case studied in Fig.~\ref{fig1}. For comparison purposes we have 
also included the LSA predictions for an impurity in the bulk, and for a 
system with two impurities, one in the bulk and one at the surface. Clearly,
an antiferromagnetically coupled bulk impurity with $S_I>S$ is energetically 
more favorable than a surface one, suggesting that self-assembled 
nanomagnetic systems with this type of impurity will tend to accumulate the 
impurities in the bulk, and leave the surface homogeneous. The ratio of the
energy difference between bulk and surface impurities to temperature is the
controlling parameter for studies of impurity migration. We find that this
parameter depends strongly on the impurity spin $S_I$ and system geometry.
The extra computational effort for applying the LSA is small, and does not 
increase significantly if more than one impurity is present, thus opening 
the possibility to model realistic nanomagnetic systems, with hundreds
or thousands of sites, with LSA.

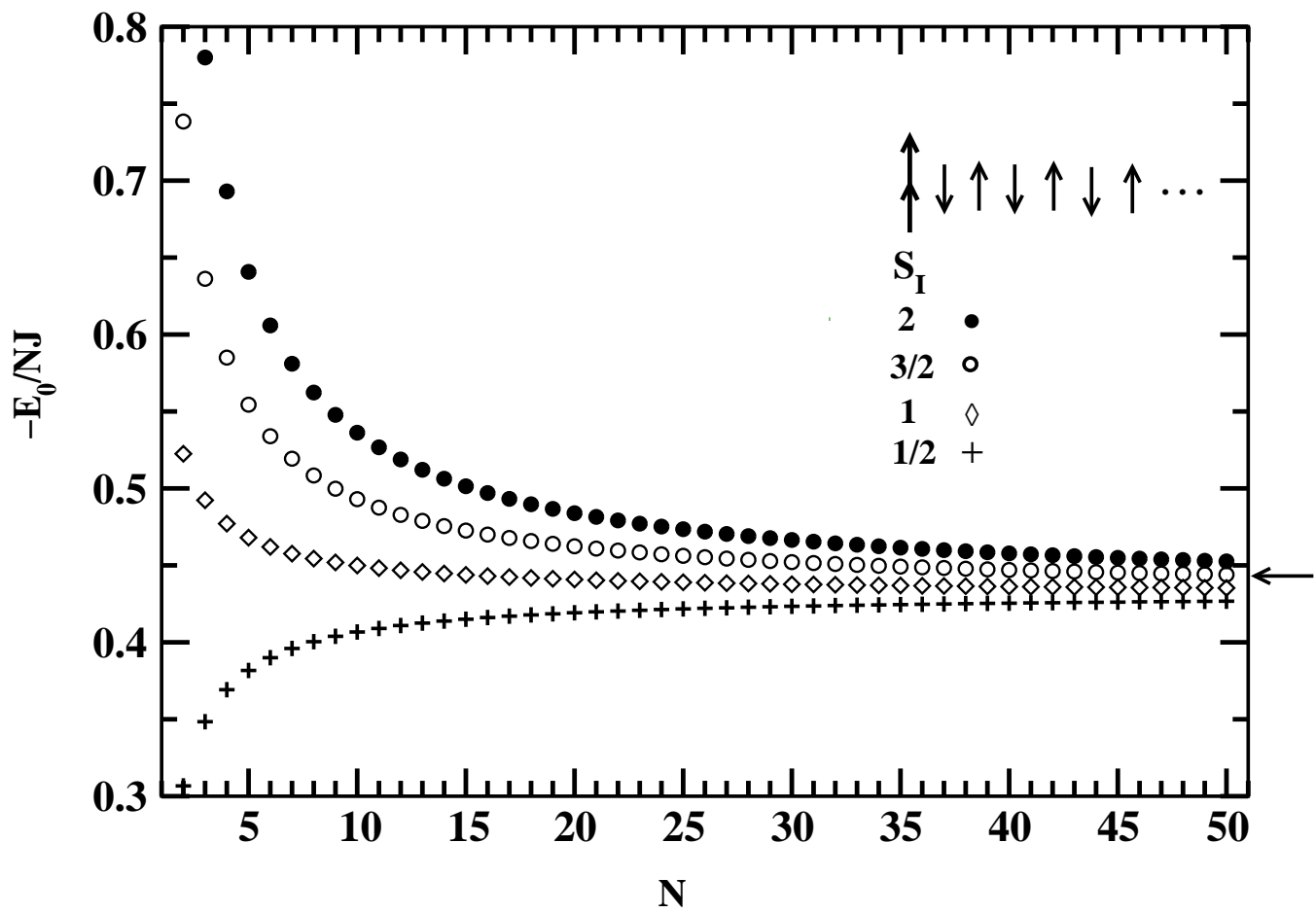
\begin{figure}
\caption{\label{fig3}Ground-state energy of an antiferromagnetic $S=1/2$
Heisenberg chain with different impurities $S_i$ at the boundary. 
Mean-field values are not included because from Fig.~\ref{fig2} we see that 
they are much inferior to LSA ones for this system. Crosses: homogeneous 
system. Diamonds: impurity spin $S_I=1$, open circles: impurity spin $S_I=3/2$,
full circles: impurity spin $S_I=2$. The arrow indicates the exact Bethe-Ansatz
result $E_0/NJ=-0.443147$.}
\end{figure}

Next, we consider an antiferromagnetic background of spin $S=1/2$ and varying 
size, with an impurity of spin $S_I>S$ at the surface. Fig.~\ref{fig3} shows
how the combined system background+impurity approaches the thermodynamic 
limit for each value of spin $S_I$. The lowest curve, representing the 
uniform system, approaches the thermodynamic limit from below, and converges
to it rather rapidly. The upper three curves represent the behaviour for 
different sizes of the impurity spin. The impurity curves approach the 
thermodynamic limit from above, at a rate that --- due to the spin-dependence 
of the prefactors of the overall $1/N$ decay --- decreases with increasing 
impurity spin. 
Physically, this behaviour can be understood by noting that for small $N$ the 
impurity greatly increases the energy of the model as compared to the uniform 
case, while, as the system size approaches infinity, the influence of the 
impurity localized at the boundary becomes negligible, and all curves converge 
to the value $E_0/JN=-0.432$, close to the value $E_0/JN=-0.443$ obtained 
from the Bethe Ansatz for the uniform system. 

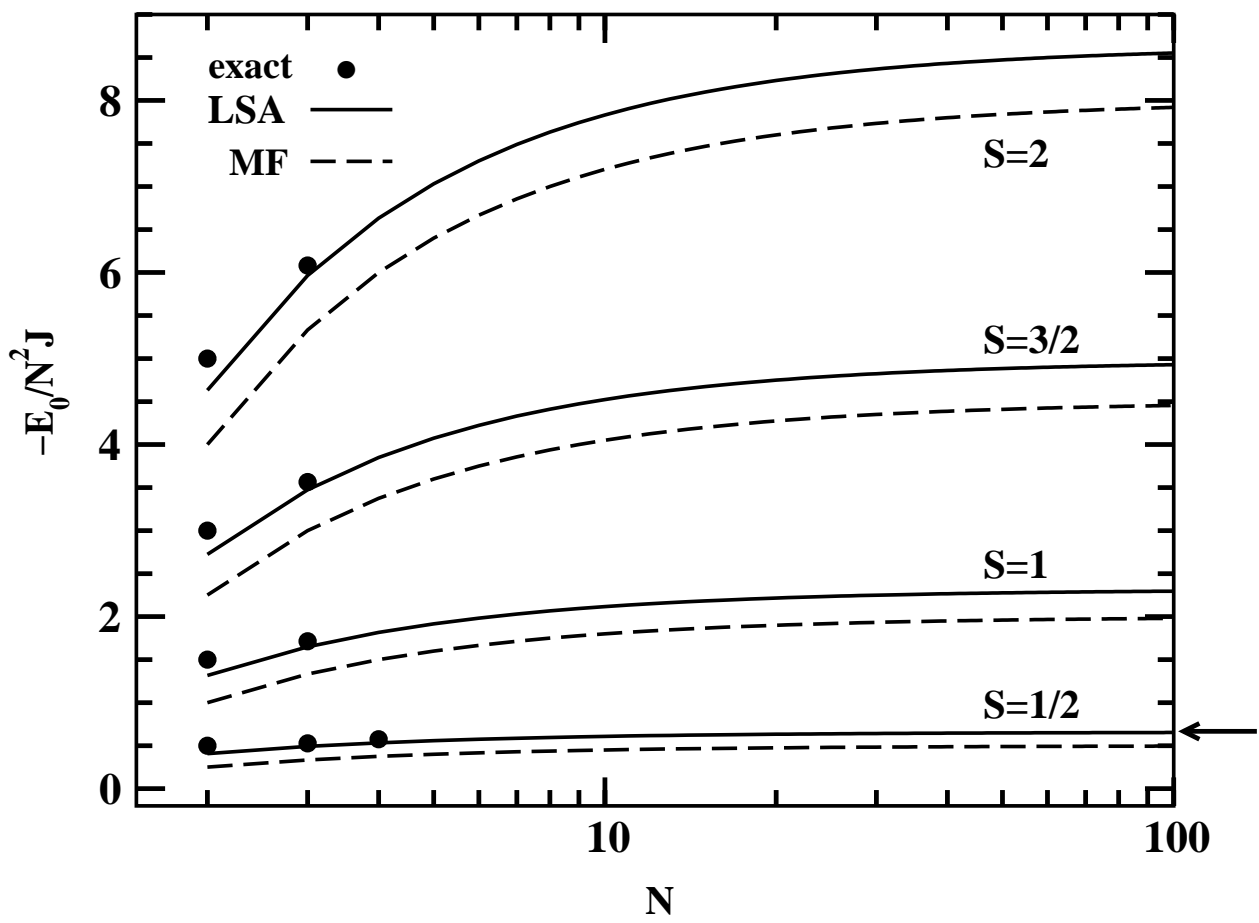
\begin{figure}
\caption{\label{fig4}Ground-state energy of finite antiferromagnetic
Heisenberg square lattices, as a function of spin and system size.
Here $N$ stands for the number of sites along the side of the square,
hence the total number of sites is $N^2$. Dashed curves: mean-field (MF)
results. Full curves: LSA results. Full circles: benchmark data obtained
by exact diagonalization of small clusters. Arrow: $N\to \infty$ limit, 
from Ref.~\onlinecite{efstratios}.}
\end{figure}

Next, we turn to the two-dimensional case. Here exact diagonalization
becomes prohibitively expensive at even smaller system sizes than in
one dimension, and 
alternative methods, such as DMRG, also encounter very significant
difficulties. In Fig.~\ref{fig4} we plot the LSA ground-state energy of
finite-size two-dimensional Heisenberg models. These are calculations
without impurities, so that the inhomogeneity arises only from the 
presence of the system boundary, which breaks translational invariance.
(As in one dimension, impurities of arbitrary size and location can be
added to these systems without a significant increase in computational cost.)
Given the LSA functional (\ref{ldafunc2d}), the generation of the LSA data 
is a very simple extension of the mean-field calculation. In spite of its
simplicity, however, this calculation illustrates an important aspect of LSA:
it has the same margin of error in higher dimensions as in one. This is rather 
untypical of many-body methods, which usually work better in certain 
dimensionalities (often $d=1$) than in others.

Finally, we investigate an effect that does not exist in one dimension, 
namely the dependence of the ground-state energy on the geometry of the 
system. In Table \ref{table1} we compare the ground-state energies of five
Heisenberg models with 100 sites, of shape $1\times 100$, $2\times 50$, 
$5\times 20$, $10\times 10$ and $2\times 5\times 10$. These systems 
represent, respectively, a one-dimensional chain of the type studied in 
Fig.~\ref{fig1}, a spin ladder with constant coupling along legs and rungs, 
a multi-legged ladder, a square lattice of the type investigated in 
Fig.~\ref{fig4}, and a representative three-dimensional structure.
The data show that: (i) for larger spins the effect of the system boundary 
becomes more pronounced. This may be relevant 
for the study of quantum corrals,\cite{corrals} and differently shaped
magnetic quantum dots. (ii) The higher-dimensional the system, the 
slower is the convergence to the thermodynamic limit: our result for the 
100 site spin $1/2$ chain differs only by 3$\%$ from the thermodynamical 
limit obtained from the Bethe-Ansatz, whereas with the same number of spins 
the energy of the 10 $\times$ 10 square lattice is about $9\%$ from the 
thermodynamical limit reported in Ref.~\onlinecite{efstratios}; a deviation 
of $3\%$ from this limit is obtained only for a 50$\times$50 lattice. 
(iii) The step from a chain to a two-legged ladder is much bigger than 
that from a two-legged ladder to a square lattice, suggesting that in 
terms of their ground-state energy even small ladders are more similar to 
extended two-dimensional systems than to chains.
(iv) Comparison of the columns $1\times 100$ and $1\times 100(2d)$ shows 
that simple extrapolation of a higher-dimensional functional to lower 
dimensionalities results in less negative energies than the correct functional.
This observation may also be relevant for {\it ab initio} calculations applying
LDA to low-dimensional systems.

In summary, our results show that the Heisenberg LSA is quantitatively 
reliable, goes significantly beyond the mean-field approximation, and allows 
to calculate the energy of systems of considerable size and complexity.
Similar results can be obtained with QMC,\cite{qmc} but at orders of magnitude 
higher computational cost. (Unlike QMC, DFT is also not limited by a
minus-sign problem.) QMC, DMRG and exact data provide benchmarks for testing
the quality of improved density functionals. On the other hand, LSA data on 
the effects of the boundary geometry, the energetic influence of impurities 
of different sizes and locations, the approach to the thermodynamic limit, and 
effects of increased dimensionality may be useful information for analysing 
real nanoscale antiferromagnetic systems, and can also serve as a guide for 
the application of alternative many-body methods to the same type of system.

\begin{acknowledgments}
This work was supported by FAPESP and CNPq. We thank F.~C.~Alcaraz 
and A.~Malvezzi for useful discussions.
\end{acknowledgments}

\newpage

\newpage \hspace*{1cm} \newpage

\newpage \hspace*{1cm} \newpage

\begin{table}
\caption{\label{table1}
Ground-state energy $-E_0/100J$ of antiferromagnetic Heisenberg models of 
fixed size $N=100$, and varying geometries and dimensionalities. For each 
dimensionality we used the proper functional,\cite{hemo} except in the column 
labeled $1\times100$(2d), which contains values obtained for the $1d$ system 
with the $2d$ functional (see discussion in main text).} 
\begin{ruledtabular}
\begin{tabular}{ccccccc}
$S$&$1\times100$& $1\times100$(2d)& $2\times50$&$5\times20$&
$10\times10$&$2\times5\times10$ \\
\hline
$1/2$ & 0.429 & 0.406 & 0.528 & 0.596 & 0.608 & 0.696 \\
$1$   & 1.353 & 1.306 & 1.796 & 2.066 & 2.116 & 2.491 \\
$3/2$ & 2.795 & 2.702 & 3.804 & 4.412 & 4.524 & 5.387 \\
$2$   & 4.687 & 4.592 & 6.552 & 7.632 & 7.832 & 9.382 
\end{tabular}
\end{ruledtabular}
\end{table}
\end{document}